# A Global Dataset Mapping the AI Innovation from Academic Research to Industrial Patents

Haixing Gong[1], Hui Zou[1,2], Xingzhou Liang[1], Shiyuan Meng[1], Pinlong Cai[1], Xingcheng Xu[1], Jingjing Qu[1*]

[1]Shanghai Artificial Intelligence Laboratory, Shanghai 200232, P. R. China.

[2]Department of Library and Information, Shanghai University, Shanghai, 200444, P. R. China.

*Corresponding author. Email: qujingjing@pjlab.org.cn

# Abstract

In the rapidly evolving field of artificial intelligence (AI), mapping innovation patterns and understanding effective technology transfer from research to applications are essential for economic growth. However, existing data infrastructures suffer from fragmentation, incomplete coverage, and insufficient evaluative capacity. Here, we present DeepInnovationAI, a comprehensive global dataset containing three structured files. DeepPatentAI.csv: Contains 2,356,204 patent records with 8 field-specific attributes. DeepDiveAI.csv: Encompasses 3,511,929 academic publications with 13 metadata fields. These two datasets leverage large language models, multilingual text analysis and dual-layer BERT classifiers to accurately identify AI-related content, while utilizing hypergraph analysis to create robust innovation metrics. Additionally, DeepCosineAI.csv: By applying semantic vector proximity analysis, this file contains 3,511,929 most relevant paper-patent pairs, each described by 3 metadata fields, to facilitate the identification of potential knowledge flows. DeepInnovationAI enables researchers, policymakers, and industry leaders to anticipate trends and identify collaboration opportunities. With extensive temporal and geographical scope, it supports detailed analysis of technological development patterns and international competition dynamics, establishing a foundation for modeling AI innovation and technology transfer processes.

## Background & Summary

Artificial Intelligence (AI) plays a pivotal role in driving the global innovation ecosystem, fostering economic growth, and enhancing technological competition[1,2]. The global AI market is projected to reach $826.7 billion by 2030, growing at a compound annual rate of 28.46% from 2024 to 2030[3]. Meanwhile, AI-related patent applications worldwide have surged, exceeding a 30% annual growth rate between 2014 and 2023, according to the World Intellectual Property Organization[4]. This rapid expansion highlights the accelerating pace of technological innovation and industrial progress. Academic papers and patents are vital for tracking AI development. Papers showcase scientific breakthroughs, while patents reflect technological applications[5,6,7]. Analyzing these two sources together offers a clearer picture of global AI innovation, revealing how discoveries evolve into practical solutions. This integrated approach also sheds light on regional innovation strengths and global competitive trends, providing valuable insights for researchers and policymakers.

However, creating comprehensive AI innovation datasets faces three significant challenges: fragmentation, limited coverage, and insufficient evaluation. First, existing databases operate in isolation. Academic platforms like Google Scholar and Dimensions focus primarily on research articles, while patent databases such as Google Patents and Derwent Innovation concentrate on technological implementations. This separation complicates efforts to connect theoretical advances with their practical applications. Most research has analyzed either papers or patents independently, with few attempts to create integrated cross source datasets[8]. Furthermore, many databases specialize exclusively in either papers or patents, leading to a disconnection that hinders the ability to observe and predict innovation through the entire process[9,10]. Additionally, restrictions in geographic coverage and timeliness have hindered global comparative analyses[11,12]. Research using patent data often focuses on single jurisdictions, such as USPTO patents[9]. Even platforms with wider coverage, like Google Patents, lack specialized AI classification systems needed for detailed analysis. Differences in language and inconsistent technical classification standards further complicate cross-regional innovation comparisons[13,14]. Most significantly, existing datasets typically lack quantitative measures of innovation, which is critical for distinguishing between incremental improvements and transformative breakthroughs, and for assessing competitive intensity. This fragmentation ultimately limits the capacity to understand the full spectrum of AI innovation dynamics. Therefore, developing a global AI dataset that integrates both papers and patents while supporting analysis of innovation trajectories and competitive dynamics would significantly advance research into science-technology collaborative innovation.

With the exponential growth of unstructured textual data, quantitative text mining has become essential in innovation research[15,16]. Text mining uses computational techniques to analyze large volumes of text, uncovering patterns and relationships that might not be visible through manual review reference. Historically, researchers have relied on keyword dictionaries to sort and study these texts[17,18]. This method involves searching for specific terms, such as "artificial intelligence" or "machine learning," within documents. While this approach is easy to implement, it has significant limitations. It often neglects the contextual nuances of word usage, leading to potential misinterpretations. In fast-evolving fields like artificial intelligence, the rapid emergence of new terminology makes fixed keyword lists insufficient, hindering the evaluation of innovation and failing to capture the dynamic nature of the discipline[18,19]. For example, Damioli et al. (2024)[11] used about 50 predefined keywords to identify AI patents, but this method may not fully capture recent advances in areas like natural language processing, potentially underestimating the extent of AI innovation. Recognizing these issues, researchers have started turning to more advanced machine learning techniques, particularly deep learning methods for natural language processing[10,17,20]. A key development in this shift is BERT (Bidirectional Encoder Representations from Transformers), a model that has transformed text analysis[17,21]. Research shows that BERT excels at tasks like classifying texts, even across multiple languages[10,20,22]. Despite these advances, current studies using BERT for innovation research face challenges. Many rely on small datasets[18], which may not represent the wide range of language found in global innovation records. Furthermore, much of the work has focused on testing how well the model performs in controlled settings, rather than applying it to massive, real-world collections of texts[17,20]. As a result, we lack thorough evidence of BERT's effectiveness when scaled up to analyze millions of papers and patents worldwide.

    This study introduces DeepInnovationAI, a comprehensive integrated dataset designed to analyze global AI technology innovation and transfer. By combining academic papers and patent records, DeepInnovationAI overcomes the limitations of

existing datasets, which often fall short in geographical coverage, time span, and technological classification. DeepInnovationAI offers several key contributions:

(1) A deep learning-based multilingual text processing system that uses advanced techniques, such as keyword matching, large language models, and dual-layer BERT classifiers. This system significantly improves the accuracy of identifying AI-related content.

(2) As the most extensive dataset of its kind, DeepInnovationAI includes 3,511,929 academic papers and 2,356,204 patent records. Covering more than 60 years, it captures data from major AI innovation countries and regions globally.

(3) By calculating text similarity between research papers and patents using KeyBERT (Keyword Extraction with BERT) and Doc2Vec (Document to Vector), we construct a cosine similarity matrix that quantifies the knowledge transfer between research and application, thereby enabling the measurement of AI innovation diffusion patterns.

(4) Novel computational framework using hypergraph analysis[15] to evaluate innovation novelty by measuring the statistical rarity of knowledge combinations, extending beyond citation metrics to objectively identify technological breakthroughs and competitive dynamics in AI evolution.

Together, these advancements position DeepInnovationAI as a valuable resource. The dataset aims to provide academic and industrial communities with a broader, more dynamic view of global AI technology innovation, supporting deeper insights into its evolution and impact.

# Methodology: Construction of the DeepInnovationAI Dataset

## Overview of the Research Framework

The DeepInnovationAI dataset follows a structured approach of "data integration-quantitative evaluation-network mapping". This dataset contains structured records globally spanning 1960-2020, comprising 3,511,929 academic papers, 2,356,204 patents, and 3,511,929 most relevant paper-patent pairs. This integration transcends traditional disciplinary barriers between academic literature and patent data, offering a comprehensive framework to study the entire innovation process, from "scientific discovery → technological invention → industrial application." The research framework is built upon three main modules (Fig. [1]):

1. **Text Acquisition and Classification (Module 1):** Implements a hierarchical framework combining IPC codes, GPT-4, and BERT classifiers to identify AI documents, generating DeepPatentAI.csv and DeepDiveAI.csv datasets.

2. **Hypergraph-Driven Innovation Quantification (Module 2):** Calculates patent novelty using probabilistic hypergraph modeling following Shi & Evans (2023)[15] advanced methodology, producing the Novelty field in both DeepPatentAI.csv and DeepDiveAI.csv files.

3. **Paper-Patent Similarity (Module 3):** Leverages KeyBERT and Doc2Vec to measure semantic associations between academic papers and patents, generating the DeepCosineAI.csv similarity matrix.

**Figure 1 goes here**

## Text Acquisition and Classification

The classification of global AI patents faces three main challenges: multilingual semantic ambiguity (manifested in the terminological ambiguity of non-English texts), dynamic technological boundaries (as seen with emerging concepts like "diffusion models"), and annotation noise (derived from strategically ambiguous expressions in patent claims). To address these challenges, a hierarchical classification framework is proposed in this study, structured as a "coarse screening - fine screening - verification" process (Fig. 2). This framework integrates keyword matching, Large Language Model (LLM) inference, and dual-layer BERT classifiers to accurately identify and categorize AI patents. The process consists of four steps:

- **Data Acquisition and Initial Filtering:** Conduct preliminary screening using International Patent Classification (IPC) codes and keywords.
- **LLM Annotation:** Enhance semantic understanding of patent texts with large language models.
- **Hierarchical Classifier Construction:** Develop and train dual-layer BERT classifiers for precise categorization.
- **Dataset Validation:** Ensure accuracy through manual review and cross-validation.

This framework complements our previously developed DeepDiveAI for academic papers by addressing the unique challenges of classifying patent data. For more details on the DeepDiveAI academic paper dataset, please refer to our earlier research[10].

**Figure 2 goes here**

**Data Acquisition and Initial Filtering**

Our research drew on the Intelligent Innovation Dataset (IIDS), a comprehensive collection of nearly 100 million patent records. This dataset was developed through a collaboration involving the Research Center for Social Intelligence at Fudan University, Boguan Innovation (Shanghai) Big Data Technology Co., Ltd., and Shanghai Artificial Intelligence Laboratory. The dataset is publicly available on the OpenDataLab platform (https://opendatalab.com/Gracie/IIDS)[23]. Its reliability, comprehensiveness, and scholarly rigor have been validated through peer review, as detailed in Wu et al. (2025)[24]. The patent corpus within IIDS was systematically extracted from the European Patent Office's official repository (https://worldwide.espacenet.com/)[25], encompassing records spanning from 1950 to the present. To ensure exceptional data quality, the IIDS underwent a multi-stage preprocessing procedure that included redundancy elimination, sophisticated missing data imputation techniques, statistical outlier detection and treatment, and meticulous expert validation procedures. It stands out for its broad patent coverage and consistent data structure. Each patent record in the IIDS includes structured fields such as unique identifiers, titles, abstracts, applicant information, and International Patent Classification (IPC) and Cooperative Patent Classification (CPC) codes, among others. For a complete description of the IIDS construction methodology and validation metrics, please refer to Wu et al. (2025)[24]. For the purposes of this research, we prioritized the patent titles and abstracts, as these sections provide the essential text needed by both human experts and automated systems to assess whether a patent relates to artificial intelligence.

To address the challenges of limited labeled data and insufficient training examples,

an "AI Keywords and International Patent Classification (IPC) Relationship Table" was constructed. This resource integrates standards from the "WIPO PATENTSCOPE Artificial Intelligence Index"[26] and the "Relationship Table between the Classification of Strategic Emerging Industries and the International Patent Classification (2021) (Trial)"[27]. The "WIPO PATENTSCOPE Artificial Intelligence Index" (https://www.wipo.int/tech_trends/zh/artificial_intelligence/patentscope.html)[26] was selected as our principal reference due to its authoritative status and comprehensiveness as a resource published by the World Intellectual Property Organization (WIPO). This index systematically organizes IPC codes and technical keywords related to artificial intelligence (AI), encompassing AI techniques, functional applications, application domains, and general AI terminology. Its comprehensive scope makes it the most relevant and reliable source for identifying AI-related patents. The "Relationship Table (2021)" (https://www.cnipa.gov.cn/art/2021/2/10/art_75_156716.html)[27] served as a supplementary reference to map correspondences between strategic emerging industries and IPC codes, thereby enhancing our understanding of AI applications in emerging sectors. Although published in 2021, this source was utilized as an auxiliary tool, with the core analysis anchored on the more current "WIPO PATENTSCOPE Artificial Intelligence Index." The resulting table includes technical terms and their corresponding IPC classifications across core AI domains such as machine learning, deep learning, and natural language processing.

Using this "AI Keywords and International Patent Classification (IPC) Relationship Table", we conducted an initial sorting of a patent dataset comprising nearly 100 million records into two preliminary categories: patents likely related to artificial intelligence (coarse classification: AI) and those unlikely to be related (coarse classification: Non-AI). This preliminary classification established a foundational starting point for identifying AI-relevant innovations. However, traditional keyword-based methods, such as the "Keywords + IPC" approach, have inherent limitations, including challenges in capturing contextual semantic information and adapting to emerging technical terminology. To mitigate these issues, we employed the "Keywords + IPC" methodology solely as an initial filtering step within a hierarchical classification framework. This coarse-grained screening was critical for managing the massive data volume, as exclusive reliance on Large Language Models (LLMs) or manual annotation would have been computationally and financially impractical. Following this initial step, deeper semantic analyses were performed using an LLM (GPT-4) and dual-layer BERT classifiers. These advanced models analyzed patent titles and abstracts to accurately identify patents substantially involving AI technologies, even when conventional keywords were absent. In essence, the primary role of the "Keywords + IPC" filtering was to establish an initial training dataset for the subsequent BERT model, achieving a balance between computational efficiency and classification accuracy.

**LLM Annotation**

Following the completion of the initial screening process, which mapped AI-related keywords and patent classifications, we used a large language model (LLM) to automatically determine whether patent texts were relevant to AI. For this task, the GPT-4 model, developed by OpenAI was employed as the annotation tool. This model, pre-trained on vast datasets, has demonstrated outstanding performance across multiple evaluation benchmarks. This approach allowed us to efficiently analyze patent documents and identify those related to AI innovation. The annotation process unfolded in three clear steps. First, we combined carefully designed prompts with the patent text, including its title and abstract. Next, we submitted this combined input to the GPT-4 model in a single interaction. Finally, the model generated a binary classification output ("yes" or "no") to determine whether the patent was related to AI.

Within the hierarchical classification framework, the LLM annotation was integrated with the "keyword + IPC" filtering method and the subsequently introduced few-shot trained coarse classification BERT model (BERT_outer). These components worked together to achieve the clear objective of constructing a large-scale initial annotated dataset. This dataset was intended to quickly provide initial binary labels ("yes" or "no") indicating whether patent texts were relevant to AI. It is important to note that this phase was designed exclusively to supply large-scale training samples for a downstream final classification model (BERT_inner), rather than to produce definitive classifications for the entire dataset. Consequently, confidence scores were neither designed nor generated during this initial annotation stage; instead, efforts concentrated on efficiently and accurately producing preliminary annotations to meet the demands of subsequent model training. To ensure annotation accuracy, specialized prompt templates were developed to guide the model's attention toward core AI technologies and application domains, including, but not limited to, machine learning, deep learning, neural networks, natural language processing, large language models, and computer vision. This guidance mechanism guaranteed the precision of annotation results and maintained strong alignment with the research objectives. Detailed descriptions of the prompt templates are provided in the Supplementary Information (Note S1). Overall, the results of the LLM annotation played a critical role in the study by generating an initial labeled sample set, which was subsequently utilized to train and evaluate the next-stage BERT classifier. Through this process, a seamless connection was established between initial annotation and subsequent model refinement, thereby laying a foundation for the entire classification framework.

**Hierarchical Classifier Construction**

To address the challenges of processing large-scale patent data while balancing classification accuracy and computational efficiency, a dual-layer classifier architecture was developed. This architecture comprises an outer coarse classifier (BERT_outer) and an inner fine-grained classifier (BERT_inner). A primary challenge was constructing a large-scale (i.e., million-sample), high-quality training dataset for BERT_inner. Given the prohibitive cost of using GPT-4 for extensive individual patent annotation, a multi-stage data processing and annotation workflow was implemented. This workflow utilized "Keywords + IPC" filtering, LLM-assisted annotation, and BERT_outer screening to progressively refine a high-quality labeled dataset suitable for BERT_inner training.

a) Outer Classifier (BERT_outer)

BERT_outer, the outer classifier, was developed using an "LLM screening and small-scale training" strategy to function as a coarse filter. An initial subset of approximately 60,000 labeled samples was generated via Keyword+IPC filtering and preliminary GPT-4 annotation. Drawing on relevant research[18] that trained BERT models using several thousand samples, the BERT_outer classifier was trained on this scale of dataset. Experimental evaluation indicated that this model could identify non-AI patents with high accuracy when applied to large-scale unlabeled data, but exhibited lower precision for AI patents. This performance profile makes BERT_outer suitable for coarse screening: it efficiently filters large volumes of definitive non-AI patents, substantially reducing the data requiring LLM annotation and associated costs, while concurrently supplying reliable non-AI negative samples for BERT_inner training.

b) Inner Classifier (BERT_inner)

To mitigate the precision limitations of BERT_outer in AI patent identification, the inner fine-grained classifier, BERT_inner, was trained on a large-scale (1 million samples), cross-validated, high-quality, balanced dataset engineered for high-precision final classification. This dataset, comprising 500,000 AI and 500,000 non-AI samples, was constructed using multi-level cross-validation and balanced sampling strategies.

- AI samples originated from: (1) patents initially filtered by "Keywords + IPC" and confirmed as AI GPT-4; (2) patents initially classified as non-AI but subsequently reclassified as AI after GPT-4 review and BERT_outer prediction, thereby incorporating samples potentially overlooked by traditional methods.

- Non-AI samples originated from: (1) patents confirmed as non-AI by both "Keywords + IPC" initial screening and BERT_outer; (2) patent records initially classified as AI but subsequently determined to be non-AI by both

GPT-4 review and BERT_outer prediction, enhancing sample robustness.

To ensure balance and representativeness across technical domains, stratified sampling based on patent IPC classification codes was conducted, maintaining balanced distribution of AI and non-AI samples across major technical categories. The final 1-million-sample balanced dataset was partitioned into training (800,000), validation (100,000), and test (100,000) sets, adhering to an 8:1:1 ratio. This strategy, combining multi-source validation with balanced stratified sampling, achieved an effective balance between data construction cost, annotation accuracy, and domain representativeness, establishing a solid data foundation for training the BERT_inner classifier. In the final classification stage, BERT_inner outputs binary predictions (AI/non-AI) and associated probability scores via its Softmax layer, indicating classification confidence. Precision-Recall (PR) curves were generated using these probabilities, and the Area Under the PR Curve (AUPRC) was calculated to comprehensively evaluate BERT_inner's performance across various decision thresholds.

**Dataset Validation**

A systematic validation process was implemented for the DeepInnovationAI dataset and associated models. First, BERT_inner model performance was assessed using the independent, stratified test set (100,000 samples). Second, to further evaluate model accuracy and generalization, five domain experts manually annotated an independent evaluation set of 6,000 samples. This expert-annotated set—distinct from the training, validation, and test partitions—provides a high-quality benchmark using previously unseen data. It will be open-sourced alongside the DeepInnovationAI dataset, providing a valuable reference for future research in AI patent classification. The integration of domain-specific terminology mapping, the dual-layer classification architecture, and deep semantic analysis established a comprehensive data processing framework capable of effectively addressing the challenges inherent in large-scale patent classification.

**Hypergraph-Driven Innovation Quantification**

Measuring the novelty of technological innovations is fundamental for identifying true breakthroughs in science and technology. Traditional citation-based metrics retrospectively track impact but fail to capture the surprising combinations that underlie transformative discoveries[28,29]. Shi & Evans (2023)[15] demonstrated that unexpected configurations of knowledge elements—those least predictable by existing patterns—are strong predictors of outsized scientific impact. Building on this understanding, our approach quantifies novelty by first deconstructing documents (academic papers and patents) into constituent technological elements. We then analyze the relationships between these elements using a hypergraph structure to evaluate their innovative potential. This method allows us to incorporate novelty indicators derived directly from the hypergraph's properties, calculated through a clear, multi-step process.

In the constructed hypergraph, where node set represents technical elements, while hyperedge set denotes high-order combinatorial relationships among these elements. Unlike traditional graphs that are limited to depicting pairwise relationships, a single hyperedge in our model can link numerous elements simultaneously, thus directly representing multivariate technological combinations. Specifically, each node corresponds to a technical keyword (such as "machine learning," "neural networks," etc.), while each hyperedge corresponds to a document, connecting all technological elements appearing within it. The advantage of this representation lies in its ability to directly capture multivariate combination patterns among technological elements, rather than being limited to pairwise relationships. The novelty of these technological combinations is then quantified through the following three steps:

- **Feature extraction:** The KeyBERT method is employed to extract key technical features from the literature. As a BERT-based language model, KeyBERT's semantic understanding enables accurate and reliable identification of important technological terms, ensuring robust foundations for subsequent analysis.

- **Hypergraph embedding and high-dimensional space modeling:** A hypergraph embedding model is adopted to map technical nodes into high-dimensional semantic space, with its core being the mixed-membership stochastic block model (MMSB). Each node $i$ is assigned a multi-dimensional vector $\theta_i$, where $\theta_{i,d}$ represents the probability distribution of node $i$ at the d-th latent dimension. These dimensions $d$ are not predefined disciplinary labels but rather "latent topics" optimally learned from data by the model, reflecting different aspects of technological elements in semantic space.

- **Novelty quantification:** Building upon these embeddings, the propensity for combination $\lambda_h$ is defined as:

$$\lambda_h = \sum_d \prod_{i \in h} \theta_{i,d} \times \prod_{i \in h} r_i \tag{1}$$

where $\prod_{i \in h} \theta_{i,d}$ measures the joint probability of nodes in combination $h$ on latent dimension $d$, representing the linguistic tightness of technical elements in the linguistic space. The term $\prod_{i \in h} r_i$ denotes the product of cognitive salience for each node in the combination, where $r_i$ is obtained through normalizing its historical occurrence frequency. The occurrence of each combination is modeled as a Poisson process:

$$X_h \sim Possion(\lambda_h) \quad (2)$$

Based on this, the novelty of technical combination $h$ is measured through negative log-likelihood:

$$Novelty(h) = -log \left( \sum_d \prod_{i \in h} \theta_{i,d} \times \prod_{i \in h} r_i \right) \quad (3)$$

The lower the propensity $\lambda_h$, the smaller the expected probability of its historical occurrence and thus higher novelty.

This framework integrates the semantic understanding capabilities of deep learning, the high-order modeling capacity of hypergraphs, and the quantification methods of probability theory, providing a systematic solution for evaluating technological innovation. For detailed mathematical derivations and implementation specifics, we refer the reader to Shi & Evans (2023)[15].

## Paper-Patent Similarity Module

To characterize the potential knowledge associations between fundamental research and technological innovation in the AI field, this study proposes a highly efficient method for detecting topical connections between academic papers and patents.

In contrast to traditional approaches that calculate pairwise similarities across entire datasets, this method focuses on precisely matching each academic paper with its most relevant subsequent patents. To ensure the temporal validity of knowledge flow, for each paper (published in year $y_i$), only patents applied for after paper publication are considered as matching candidates ($y_j \geq y_i$), thereby effectively eliminating the possibility of reverse influence. When processing large-scale data, cosine similarity is employed as the text similarity metric based on three primary considerations: first, its mathematical robustness in semantic space effectively captures topical similarities; second, its computational efficiency and scalability are well-suited to processing large-scale corpora; and third, existing research[30,31] has validated its practical value in analyzing technological innovation associations. The core process of this method comprises the following three steps:

- Semantic feature extraction: The process began by using the KeyBERT model to extract key technological terms from both scholarly articles and patent texts, establishing an initial feature space. Next, the Doc2Vec model was applied to map the texts into a high-dimensional vector space, capturing document-level semantic information.

- Cosine similarity calculation: Based on the extracted feature vectors, the association strength between each pair of scholarly articles and patent documents was calculated using cosine similarity:

$$Similarity(d_i, p_j) = \frac{\vec{v}_i \cdot \vec{v}_j}{\|\vec{v}_i\|\|\vec{v}_j\|} \tag{4}$$

where $\vec{v}_i$ represents the feature vector for paper $d_i$ and $\vec{v}_j$ denotes the feature vector of patent $p_j$.

- Matching to the most relevant patent: For each paper $d_i$ (published in year $y_i$), only patents $p_j$ with application year $y_j \geq y_i$ are considered. Among these candidate patents, the one exhibiting the highest semantic similarity is deemed the most relevant, denoted as $p_{j^*}$, where:

$$j^* = \arg\max_{j, y_j \geq y_i} Similarity(d_i, p_j) \qquad (5)$$

$$BestMatch_i = p_{j^*} \qquad (6)$$

**Data Records**

The DeepInnovationAI dataset[32] has been uploaded to Figshare and is publicly accessible via https://doi.org/10.6084/m9.figshare.28578947. The dataset features a modular structure with three core files: patent data (DeepPatentAI.csv), academic papers (DeepDiveAI.csv), and paper-patent similarity (DeepCosineAI.csv).

DeepPatentAI.csv contains 2,356,204 patent records with fields detailed in Table 1, including basic patent information, technical classifications, extracted keywords, and innovation metrics calculated using hypergraph models. Similarly, DeepDiveAI.csv encompasses 3,511,929 academic papers, featuring metadata, keyword features, and innovation indicators as outlined in Table 2. Additionally, DeepCosineAI.csv contains 3,511,929 most relevant paper-patent pairs with fields detailed in Table 3, including Paper_ID, Patent_ID, and Similarity.

**Table 1** Description of DeepPatentAI dataset.

| Features | Data type | Description |
| --- | --- | --- |
| ID | Int | Sequential identifier ranked from 1 to 2,356,204 |
| PN | String | The unique patent number |
| IPC | String | The International Patent Classification (IPC) code |
| Title | String | The title of the patent |
| Abstract | String | The abstract summarizing the patent |
| Year | Int | The year in which the patent application was filed |
| Keywords | String | A JSON-formatted list of keywords |
| Novelty | Float | A numerical indicator quantifying the patent Innovation |

**Table 2** Description of DeepDiveAI dataset. Details can be found in Zhou et al. (2024)[10]

| Features | Data type | Description |
|---|---|---|
| ID | Int | Sequential identifier ranked from 1 to 3,511,929 |
| Title | String | The title of the paper |
| Source | String | The title of the journal or conference proceedings |
| Abstract | String | The abstract summarizing the paper |
| Name | String | Names of authors who contributed to the paper |
| Year | Int | Paper publication year |
| Cited | Int | Number of times the paper has been cited |
| Keywords | String | A JSON-formatted list of keywords |
| Doi | String | Unique identifier that provides a link for online access |
| Language | String | Language of publication |
| Type | String | Type of papers |
| Affiliations | String | Detailed affiliation information |
| Novelty | Float | A numerical indicator quantifying the paper Innovation |

Table 3 Description of DeepCosineAI dataset.

| Features | Data type | Description |
|---|---|---|
| Paper_ID | Int | Sequential identifier ranked from 1 to 3,511,929 |
| Patent_ID | Int | Identifier of the most relevant patent corresponding to each paper |
| Cosine similarity | Float | Cosine similarity score between the paper and its most relevant patent |

## Technical Validation

This section presents a systematic validation of the DeepPatentAI dataset and the efficacy of its construction methodology, implemented within the DeepInnovationAI framework. Experiments were conducted on a high-quality sample set of one million entries (500,000 AI-related and 500,000 non-AI-related), which was processed via a multi-stage classification strategy involving "Keywords + IPC" filtering, LLM annotation, BERT_outer pre-screening. Standard metrics, including Accuracy, Recall, Precision, F1 score, and Area Under the Precision-Recall Curve (AUPRC), were utilized for performance evaluation. Furthermore, to rigorously assess the robustness and generalization capabilities of the processing pipeline, an independent evaluation set comprising 6,000 entries annotated by six domain experts was established as an external benchmark.

**Evaluation Metrics**

We implemented a comprehensive evaluation framework based on confusion matrix analysis[21] to assess model performance. Recognizing the potential class imbalance inherent in patent classification, we utilized multiple complementary metrics including accuracy, precision, recall, and F1 score. These metrics are defined as:

$$Accuracy = \frac{TP + TN}{TP + TN + FP + FN} \tag{7}$$

$$Precision = \frac{TP}{TP + FP} \qquad (8)$$

$$Recall = \frac{TP}{TP + FN} \qquad (9)$$

$$F1 = 2 \times \frac{Precision \times Recall}{Precision + Recall} \qquad (10)$$

where $TP$, $TN$, $FP$, and $FN$ represent True Positive, True Negative, False Positive, and False Negative classifications, respectively.

To gain deeper insight, we also employed the precision-recall curve (PR curve), which illustrates the balance between precision and recall as the model's decision threshold—the cutoff for classifying a patent as AI-related—varies:

$$PR(\theta) = (P(\theta), R(\theta)), \theta \in [0,1] \qquad (11)$$

where $\theta$ represents the decision threshold, and $P(\theta)$ and $R(\theta)$ denote the precision and recall at that threshold, respectively. The area under the precision-recall curve (AUPRC) is defined as:

$$AUPRC = \int_0^1 P(R)dR \qquad (12)$$

Typically, superior model performance is indicated by a PR curve that approaches the upper right corner of the graph, with the AUPRC value nearing 1. The PR curve is particularly valuable because it emphasizes the model's ability to identify AI-related patents without being swayed by the abundance of non-AI patents. This focus makes it well-suited for datasets where one category significantly outweighs the other, ensuring a robust evaluation of the model's effectiveness[22].

**Evaluation of Training Dataset Construction Methodology**

A primary challenge in this research was the development of a high-quality, large-scale training dataset for the final AI patent classifier, BERT_inner. This dataset construction employed a multi-stage classification methodology, comprising "Keywords + IPC" coarse filtering, LLM (GPT-4) annotation, and BERT_outer pre-screening. To systematically validate the effectiveness of these three stages, we utilized an independent evaluation set of 6,000 expert-annotated entries distinct from the training, validation, and test sets, thereby ensuring objective assessment. Table [4] presents the detailed performance metrics of each methodological stage evaluated on this independent set.

- "Keywords + IPC" filtering: As the initial screening stage, this method achieved an overall accuracy of 0.877. For the AI category, it yielded a precision of 0.801, recall of 0.905, and an F1 score of 0.856; for the Non-AI category, precision was 0.931, recall was 0.855, and the F1 score was 0.891.
- LLM annotation (GPT-4): Leveraging the semantic understanding capabilities of large language models, classification performance improved significantly, with overall accuracy reaching 0.948. For the AI category, precision, recall, and F1 scores increased to 0.951, 0.919, and 0.935, respectively; for the Non-AI category, the corresponding metrics were 0.946, 0.968, and 0.957. Notably, owing to the high cost associated with LLM, this stage was primarily employed to generate high-quality training data rather than for direct classification of the entire dataset.
- BERT_outer: As a lightweight pre-screener, BERT_outer provided a balance between efficiency and classification accuracy, achieving an overall accuracy of 0.845. For the AI category, its precision was 0.774, recall was 0.874, and the F1 score was 0.821; for the Non-AI category, precision was 0.905, recall was 0.826, and the F1 score was 0.864. The primary utility of BERT_outer stems from its ability to efficiently identify a large volume of clearly Non-AI patents, thereby significantly reducing the number of samples requiring subsequent LLM processing and lowering overall costs.

To further elucidate the complementarity and redundancy among the methods, an overlap analysis was performed on their results, as illustrated in Fig. 3.

- Method overlap proportions: The overlap rate between "Keywords + IPC" and LLM results was 86.48%; between LLM and BERT_outer, it was 83.57%; and between "Keywords + IPC" and BERT_outer, it was 89.82%. The three-way overlap among all methods was 79.9%, indicating substantial consensus. The non-overlapping portions suggest that each method also possesses unique identification capabilities. This pattern of high but imperfect overlap supports the rationale for employing a multi-method collaborative strategy.

- Accuracy analysis in overlap regions: Classification accuracy in the overlap regions between different methods was generally higher than their individual accuracies. Specifically, the accuracy within the pairwise overlap between "Keywords + IPC" and LLM was 97.7%; between "Keywords + IPC" and BERT_outer, 90.2%; and between LLM and BERT_outer, 97.5%. Accuracy within the three-way overlap region reached 98.0%. These results suggest that consensus judgments across multiple methods yield significantly higher reliability, further underscoring the benefit of multi-method collaboration.

**Table 4** Comparison of performance metrics for the individual stages ("Keywords + IPC", LLM, BERT_outer) within the multi-stage AI patent classification preprocessing methodology, evaluated on the expert-annotated dataset.

| Process | Overall Accuracy | Class | Precision | Recall | F1 Score |
| --- | --- | --- | --- | --- | --- |
| Keywords + IPC | 0.877 | Non-AI | 0.931 | 0.855 | 0.891 |
|  |  | AI | 0.801 | 0.905 | 0.856 |
| LLM | 0.948 | Non-AI | 0.946 | 0.968 | 0.957 |
|  |  | AI | 0.951 | 0.919 | 0.935 |
| BERT_outer | 0.845 | Non-AI | 0.905 | 0.826 | 0.864 |
|  |  | AI | 0.774 | 0.874 | 0.821 |

**Figure 3 goes here**

**Dataset Split**

To support robust model training and dependable evaluation, the stratified random sampling method based on patent IPC classification codes was adopted to partition the 1 million labeled patent records, maintaining balanced distributions of AI and non-AI samples across major technical categories. Specifically, the dataset was divided into training Set, validation set, and test set in a ratio of 8:1:1, comprising 800,000, 100,000, and 100,000 records respectively. This approach provided ample data for training while maintaining sufficient samples for validation and testing.

**Model and Training Details**

The final classifier, designated BERT_inner, utilizes a BERT-based deep learning architecture for patent classification. Model training was conducted on a high-performance computing cluster equipped with eight NVIDIA A100-SXM4-80GB GPUs. Patent titles and abstracts served as the input features. Following standardized preprocessing, these texts were converted into token sequences with a maximum length of 512 tokens. The model's foundation layer incorporates the pre-trained bert-base-uncased model ([https://huggingface.co/google-bert/bert-base-uncased](https://huggingface.co/google-bert/bert-base-uncased)) to process patent texts. The output layer is configured as a binary classifier, employing the Softmax function to generate probability distributions indicating whether a patent pertains to AI. Multi-GPU parallel computing techniques facilitated the efficient utilization of computational resources during training. During data preprocessing, standardization and dynamic padding methods were implemented to maintain data quality and consistency.

Model optimization utilized the AdamW optimizer with a 2e-5 initial learning rate and 0.01 weight decay coefficient, paired with linear learning rate scheduling. This approach enhanced convergence performance while minimizing overfitting. Mixed precision training reduced training time by 40% without performance degradation. We implemented a mini-batch strategy (batch size of 16) and an early stopping mechanism that terminated training after three consecutive rounds without validation improvement. The model achieved optimal performance in the second training round with a validation accuracy of 0.977. This result indicated effective training and convergence properties, while the early stopping mechanism successfully prevented overfitting.

**Evaluation Results**

The BERT_inner model demonstrated strong classification performance across all data splits (Table 5 and Fig. 4). On the training set (N = 800,000 samples), the model achieved an overall accuracy of 0.984, with precision values for non-AI and AI patents of 0.989 and 0.979, respectively, recall values of 0.981 and 0.988, and F1-scores of 0.985 and 0.983. These results indicate a high degree of model fit to the training data. Performance on the validation set (N = 100,000 samples) remained robust, showing only a slight decrease compared to the training set; the overall accuracy was 0.977, overall precision was 0.977, and the AI class recall value was 0.981. On the test set (N = 100,000 samples), despite a slight decrease in performance, the overall accuracy was 0.975. For the AI class, precision and recall values were 0.963 and 0.974, respectively, yielding an F1-score of 0.968. Evaluation of the precision-recall (PR) curve indicated strong performance, yielding an Area Under the PR Curve (AUPRC) value of 0.996. This result further supports the model's stable performance across various decision thresholds. The consistent performance across the three datasets, reflected in the overall accuracies of 0.984 (training), 0.977 (validation), and 0.975 (test), demonstrates the model's strong generalization capability. Particularly noteworthy is the high recall value (0.974) for the AI class on the test set, indicating the model's effectiveness in identifying diverse AI patent expressions; this capability is crucial for constructing a comprehensive AI patent dataset.

To rigorously assess the model's generalization capability on unseen data, the model was tested on an independent evaluation set (N = 6,000 samples) annotated by six domain experts. As shown in the final two rows of Table 5, the model exhibited robust performance on this independent evaluation set, achieving an overall accuracy of 0.908. Notably, a recall value of 0.992 was achieved for the AI patent identification task, indicating that most AI patents were accurately identified. Compared to existing research, DeepPatentAI demonstrates clear advantages. It surpasses deep learning approaches based on convolutional neural networks and word vector embeddings (0.840 accuracy)[33], as well as random forests and support vector machines (F1 scores of 0.90 and 0.85 respectively)[18]. The model also greatly exceeds traditional keyword-based classification methods (F1 score of 0.59)[34]. These results confirm the effectiveness of the DeepPatentAI framework, which not only performed excellently in standardized testing environments but also maintained stable performance in practical applications. This provides a reliable foundation for building high-quality AI patent datasets.

**Figure 4 goes here**

**Table 5** Evaluation results of classification performance of DeepPatentAI on various datasets.

| Dataset | Overall Accuracy | Class | Precision | Recall | F1 Score |
| --- | --- | --- | --- | --- | --- |
| Training Set | 0.984 | Non-AI | 0.989 | 0.981 | 0.985 |
|  |  | AI | 0.979 | 0.988 | 0.983 |
| Validation Set | 0.977 | Non-AI | 0.977 | 0.972 | 0.975 |
|  |  | AI | 0.977 | 0.981 | 0.979 |
| Test Set | 0.975 | Non-AI | 0.983 | 0.975 | 0.979 |
| | | AI | 0.963 | 0.974 | 0.968 |
| Manually Annotated Set | 0.908 | Non-AI | 0.993 | 0.837 | 0.915 |
| | | AI | 0.807 | 0.992 | 0.900 |

## Usage Notes

The DeepInnovationAI dataset facilitates comprehensive analysis of the global AI innovation landscape. The following usage notes offer guidance for the effective utilization of the data:

- **Data access and processing:** The dataset comprises three core CSV files (DeepPatentAI.csv, DeepDiveAI.csv, and DeepCosineAI.csv), all encoded in UTF-8. Standard data analysis tools, such as Python (utilizing the pandas and numpy libraries) or R, can be employed for processing. For handling larger datasets or more complex computations, distributed computing frameworks such as Apache Spark or Dask are recommended. Some fields containing technical terms are stored in JSON format; these can be readily converted into usable data structures, for instance, by employing Python's json module. The 'Novelty' field contains a numerical score, wherein higher values signify greater degrees of innovation.

- **Technology transfer:** The DeepCosineAI.csv file contains similarity matrices designed to facilitate the identification of potential knowledge flows between academic research and patents. It is important to note that the semantic similarity metrics between papers and patents provided in this dataset should be regarded as foundational tools for identifying potential knowledge flow channels, rather than as definitive assertions regarding actual knowledge transfer processes. Researchers can undertake more in-depth investigations into knowledge transfer mechanisms by integrating this baseline dataset with various supplementary indicators, such as patent citation networks, inventor-author matching, and institutional affiliations.

- **Computational resource requirements:** For effective utilization of the dataset, certain computational resources are recommended. For basic analysis, a computing environment possessing at least 8 CPU cores and 32GB of RAM is recommended. More demanding tasks, including additional natural language processing or the construction of machine learning models, would derive greater benefit from high-performance computing environments, particularly those equipped with GPUs.

- **Application scenarios:** The dataset supports diverse analyses of AI innovation, including tracking the evolution of AI technical themes, examining development patterns across time and geography, and investigating global innovation network structures. Detailed examples of such applications are provided in the Supplementary Information (Note S2, Figs. S1, S2 and S3).

- **Data updates and maintenance:** The current version of the dataset incorporates data up to 2020. It will be periodically updated to include newly published academic papers and patent records, with the latest versions accessible via the Figshare link provided.

- **Citation specification:** When utilizing this dataset for academic research, please adhere to the citation standards provided in the dataset metadata for referencing this resource and related papers.

Adherence to these guidelines will facilitate researchers in efficiently accessing and processing the DeepInnovationAI dataset. For additional technical support, the development team may be contacted via the project homepage or its GitHub repository.

## Code availability

The computational framework in this study was implemented using Python, with the BERT model developed using PyTorch. Codes in this study are available at https://github.com/Haixing99/DeepInnovationAI.

## Acknowledgements

This work was supported by the National Key Research and Development Program

of China under Grant 2022ZD0116205. We also gratefully acknowledge the computational resources provided by the Shanghai Artificial Intelligence Laboratory.

## Author contributions

H.G. led the research design, system architecture development, and implementation of the BERT model, coordinated overall project workflow, and drafted the manuscript. H.Z. managed data acquisition and integration of multilingual patent sources. X.L. and S.M. were responsible for data preprocessing, anomaly detection, and developing the coarse-grained classifier. P.C. and X.X. contributed to the technical validation of the dataset. J.Q. supervised the project, proposed the original research concept, coordinated resources across teams, and provided critical revisions to the manuscript. All authors reviewed and approved the final version of the manuscript.

## Competing interests

The authors declare no competing interests.

## Additional information

Correspondence and requests for materials should be addressed to J.Q.

## Figure legends

**Fig. 1** Architectural features of the DeepInnovationAI dataset.

**Fig. 2** Schematic diagram of the global AI patent dataset construction process.

**Fig. 3** Performance evaluation and comparative analysis of AI patent classification preprocessing methods ("Keywords + IPC", LLM, BERT_outer) using an independently expert-annotated evaluation set. **a** Comparison of accuracy across the different methods; **b** Analysis of the overlap matrix between method outputs; **c** Assessment of accuracy within distinct overlap regions; and **d** Visualization illustrating the relationship between method overlap rate and classification accuracy.

**Fig. 4** Comprehensive performance evaluation of DeepPatentAI. **a-b** Performance radar charts for Non-AI and AI classes; **c-d** PR curves on the test and manually evaluated datasets.